\documentclass{article}
\usepackage{spconf,amsmath,graphicx,url,times,amssymb,acronym,xcolor,balance}
\graphicspath{{figures/}}

\usepackage{color}
\usepackage{lipsum}
\usepackage{courier}

\usepackage{flushend} 
\usepackage{numprint}
\npfourdigitnosep
\usepackage{booktabs} 
\usepackage{siunitx}  
\usepackage{cleveref} 
\creflabelformat{equation}{#2(#1)#3}
\crefname{equation}{}{}
\crefname{figure}{Figure}{}
\crefname{table}{Table}{}
\crefname{section}{Section}{}

\acrodef{STFT}{short-time Fourier transform}
\acrodef{PSD}{power spectral density}
\acrodef{RTF}{relative transfer function}
\acrodef{SNR}{signal-to-noise ratio}
\acrodef{PDF}{probability density function}
\acrodef{VAD}{voice activity detector}
\acrodef{MVDR}{minimum variance distortionless response}
\acrodef{AIR}{acoustic impulse response}
\acrodef{PESQ}{perceptual evaluation of speech quality}
\acrodef{CD}{cepstral distance}
\acrodef{WER}{word error rate}
\acrodef{SPP}{speech presence probability}
\acrodef{RNN}{recurent neural network}
\acrodef{LSTM}{long-term short-term}
\acrodef{RMSE}{root-mean-square error}
\acrodef{PCA}{principal component analysis}
\acrodef{RHO}[$\rho$]{Pearson correlation coefficient}

\newcommand{\symvec}[1]{{\mbox{\boldmath $#1$}}}

\newcommand{\ASR}[1]{SR$_\mathbf{#1}$}

\makeatletter
\def\bstctlcite{\@ifnextchar[{\@bstctlcite}{\@bstctlcite[@auxout]}}
\def\@bstctlcite[#1]#2{\@bsphack
	\@for\@citeb:=#2\do{%
		\edef\@citeb{\expandafter\@firstofone\@citeb}%
		\if@filesw\immediate\write\csname #1\endcsname{\string\citation{\@citeb}}\fi}%
	\@esphack}
\makeatother

\title{Predicting word error rate for reverberant speech}

\name{Hannes Gamper, Dimitra Emmanouilidou, Sebastian Braun, Ivan J. Tashev}
\address{Microsoft Research,\\ One Microsoft Way, Redmond, WA, USA\\
\texttt{\{hannes.gamper, diemmano, sebraun, ivantash\}@microsoft.com}}

\begin{document}

\acused{RHO}
\bstctlcite{IEEEexample:BSTcontrol}

\ninept
\maketitle

\begin{sloppy}

\begin{abstract}
Reverberation negatively impacts the performance of automatic speech recognition (ASR).
Prior work on quantifying the effect of reverberation has shown that clarity (C50), a parameter that can be estimated from the acoustic impulse response, is correlated with ASR performance.
In this paper we propose predicting ASR performance in terms of the word error rate (WER)
directly from acoustic parameters via a polynomial, sigmoidal, or neural network fit,
as well as blindly from reverberant speech samples using a convolutional neural network (CNN).
We carry out experiments on two state-of-the-art ASR models and a large set of acoustic impulse responses (AIRs).
The results confirm C50 and C80 to be highly correlated with WER, allowing WER to be predicted with the proposed fitting approaches.
The proposed non-intrusive CNN model outperforms C50-based WER prediction, indicating that WER can be estimated blindly, i.e., directly from the reverberant speech samples without knowledge of the acoustic parameters.
\end{abstract}


\section{Introduction}
Automatic speech recognition (ASR) aims at transcribing recorded speech to text. In enclosed spaces, acoustic reflections and reverberation arrive at the receiver filtered and delayed relative to the direct propagation path, thus repeating and smearing the speech signal in time. Therefore, reverberation may negatively impact human speech intelligibility and the performance of ASR engines~\cite{Payton1994,REVERB2013}.

The REVERB challenge aimed at providing a common data set to evaluate state-of-the-art ASR models in the presence of noise and reverberation~\cite{REVERB2013}. The challenge results indicate that while the proposed ASR approaches varied substantially in terms of features, model architecture, and performance, they exhibited similar behaviour with respect to the relative increase or decrease of the word error rate (WER) as a function of the reverberation conditions~\cite{Kinoshita2016}. To increase robustness of ASR engines against reverberation, prior work includes denoising and dereverberation techniques~\cite{Feng2014}, improved speech features~\cite{Kim2016}, improved model architectures~\cite{Zhang2016}, and data augmentation techniques~\cite{Ravanelli2016,ko2017,Kim2017}.

With voice-enabled services being deployed in more challenging scenarios and becoming more ubiquitous, e.g., through the rise of smart home devices, it is useful to quantify the effect of reverberation on ASR performance. Assuming a linear and time-invariant system, the effect of reverberation on a sound signal is determined by the acoustic impulse response (AIR) of the reverberant environment. A convenient way to describe an AIR is by estimating acoustic parameters, including the \emph{reverberation time} (T60), \emph{direct-to-reverberation ratio} (DRR), \emph{clarity} (C50), and \emph{definition} (D50)~\cite{naylor2010speech}.
A recent study investigated the effect of these reverberation parameters on audio event classification~\cite{emmanouilidou2019}. Prior work on acoustic parameter estimation has indicated the usefulness of this parameterization in the context of ASR. Giri et al.\ showed that ASR performance in reverberant conditions could be improved by combining speech features with estimates for T60 and DRR~\cite{Giri2015}.
Fukumori et al.\ proposed predicting the performance of a hidden Markov model (HMM) based ASR system from a speech quality parameter (PESQ~\cite{ITU_T_P862}) and definition~\cite{fukumori2010}. Tsilfidis et al.\ showed that C50 and D50 can be used to predict phoneme recognition rate~\cite{Tsilfidis2013}. Parada et al.\ proposed a blind C50 estimator that is correlated with phoneme recognition~\cite{Parada2016}.

Here we study the effect of reverberation on the performance of ASR systems in terms of the word error rate (WER).
Speech recognition models convert an audio signal into a sequence of words, often via an intermediate phoneme representation~\cite{chiu2018state}, which lends itself to calculating a phoneme error rate (PER) as a performance metric.
Prior work has used PER as an evaluation criterion, as it is assumed to be directly impacted by reverberation and does not depend on a language model~\cite{Tsilfidis2013,Parada2016}. However, with the emergence of end-to-end speech recognition systems that map directly from acoustic input features to sequences of words~\cite{chiu2018state}, it may be difficult to derive a PER. Therefore, we focus on WER as the performance criterion instead. We evaluate two state-of-the-art ASR models on a clean speech corpus convolved with a large set of measured AIRs. First, we confirm that C50 and C80 remain the most important among a range of AIR parameters for predicting the WER. Next, we propose models for predicting WER from acoustic parameters, using a polynomial, sigmoidal, or neural network fit.
Finally, we propose a non-intrusive approach for predicting WER blindly from reverberant speech samples using a convolutional neural network (CNN).
The CNN model may be suitable for applications where neither the clean reference speech nor the raw AIRs are available. A light-weight, non-intrusive WER estimator could potentially be useful to derive a loss metric to train dereverberation models for ASR, or to predict WER for data without transcription to be used for unsupervised ASR (pre-)training~\cite{Schneider_2019}.

\section{Data corpus and metrics}
\label{sec:corpus}
To train and evaluate the proposed methods for WER prediction, we generated a large and diverse corpus of reverberated speech. Clean speech recordings were taken from the LibriSpeech ASR Corpus~\cite{LIBRISPEECH_2015}, a corpus containing speech from approximately \numprint{8000} public domain audio books, sampled at 16 kHz. We used the 100-hour training corpus for training the ASR models, and the test set for evaluation. To simulate reverberant speech, we compiled a large corpus of measured acoustic impulse responses (AIRs) from proprietary as well as publicly available data sets, in an effort to cover a wide range of acoustic conditions: the Aachen Impulse Response database~\cite{AachenAIR}, the Open Acoustic Impulse Response database~\cite{murphy2010openair}, the Multichannel Acoustic Reverberation Database at York~\cite{wen2006evaluation}, the PORI Concert Hall Impulse Responses~\cite{PORI_2005}, AIRs published in the SOFA format~\cite{sofa2018}, the REVERB Challenge database~\cite{REVERB}, SMARD~\cite{nielsen2014single}, the Echothief Impulse Response Library~\cite{EchoThief}, the Concert Hall Research Group database~\cite{bradley1994data}, the Real Acoustic Environments Working Group database~\cite{nakamura2000acoustical}, the QMUL Room Impulse Response Data Set~\cite{QMUL}, and The ACE Challenge Corpus~\cite{ACEchallenge}. After pruning AIRs with measurement-related issues or that were outliers otherwise, the set totalled \numprint{15167} single-channel AIRs.

The reverberant set used to test the ASR systems and evaluate the WER prediction performance of the proposed methods consisted of the LibriSpeech ``test clean'' corpus, containing \numprint{2620} utterances by 40 speakers. The clean utterances were convolved with randomly drawn AIRs from the set described above. This process was repeated multiple times to increase the size of the evaluation set.
For training and testing the WER prediction methods, the evaluation set was further split into \numprint{17280} training (TR), \numprint{2430} validation (CV), and \numprint{2196} test utterances (TE), such that the talkers and AIRs in TE were not contained in TR.


\subsection{Impulse response parameter estimation}
\label{sec:AIR}
The \emph{reverberation time} (\textbf{T60}) is the time it takes for the AIR energy to drop by 60~dB. It is estimated here using the method by Karjalainen et al.~\cite{karjalainen2002estimation,ACEchallenge}. An alternative way to estimate reverberation time is by fitting a line to the energy decay curve (EDC)~\cite{ACEchallenge}.
Here, we fit a line from the point where the EDC drops below -5~dB to where it drops below -15~dB (\textbf{T10}), -20~dB (\textbf{T15}), -25~dB (\textbf{T20}), and -35~dB (\textbf{T30}). A metric shown to correlate well with human perception of reverberance is the \emph{early decay time} (\textbf{EDT}), calculated at the point where the EDC first drops below -10~dB~\cite{Soulodre1995}.

The \emph{bass ratio} (\textbf{BR}) is the ratio between the average reverberation times at low and high frequencies~\cite{Soulodre1995}. With the reverberation time T$_f$ of an AIR filtered through an octave-band filter with center frequency $f$, the BR is given as~\cite{Soulodre1995}
\begin{equation}
    \mathrm{BR} = \frac{T_{125} + T_{250}}{T_{500} + T_{1000}}.
\end{equation}

The \emph{direct-to-reverberant ratio} (\textbf{DRR}) relates the energy of the direct path to the energy of reflected paths~\cite{ACEchallenge}.
The direct path energy is determined as the energy in a window of 2.5~ms around the maximum of the AIR, $h[n]$, while the energy outside this window is taken as the reverberant energy~\cite{ACEchallenge}:
\begin{equation} \label{eq:DRR}
    \mathrm{DRR} = 10 \log_{10}
    \left(
    \frac{\sum_{n=n_d-n_w}^{n_d+n_w}h[n]^2}
    {\sum_{n=n_d+n_w}^{\infty}h[n]^2}
    \right),
\end{equation}
where \vspace{-0.2cm}
\begin{equation} \label{eq:nd}
n_d = \underset{n}{\arg\max}~|h[n]|
\end{equation}
and $n_w$ denotes the number of samples corresponding to a window length of 2.5~ms. Note that \cref{eq:DRR} and \cref{eq:nd} are slightly modified compared to the definitions given by Eaton et al.~\cite{ACEchallenge} to operate on AIRs with unknown measurement characteristics.
Similar to DRR, \emph{clarity} is a measure for the energy ratio between early and late parts of the AIR~\cite{Soulodre1995}. It is given as
\begin{equation}
    \label{eq:C50}
    \mathrm{C}_{t} = 10 \log_{10}
    \left(
    \frac{\sum_{n=n_0}^{n_0+n_t}h[n]^2}
    {\sum_{n=n_t}^{\infty}h[n]^2}
    \right),
\end{equation}
where $n_0$ is defined as
the sample with the largest drop in the EDC, which was found to be a relatively robust measure for determining the direct path,
and $n_t$ is the number of samples corresponding to a window length $t$ in ms. With \cref{eq:C50}, we estimate \textbf{C30}, \textbf{C50}, and \textbf{C80}. Similarly, the \emph{definition}, $\mathrm{D}_t$, can be calculated as~\cite{Soulodre1995}:
\begin{equation}
    \label{eq:D50}
    \mathrm{D}_{t} =
    \left(
    \frac{\sum_{n=n_0}^{n_0+n_t}h[n]^2}
    {\sum_{n=n_0}^{\infty}h[n]^2}
    \right).
\end{equation}
Given \cref{eq:D50}, we estimate \textbf{D30}, \textbf{D50}, and \textbf{D80}.

The \emph{center time} (\textbf{Tc}) is defined as~\cite{Tsilfidis2013}:
\begin{equation}
    \mathrm{Tc} = \frac{\sum_{n=n_0}^{\infty}\frac{n-n_0}{fs}h[n]^2}{\sum_{n=n_0}^{\infty}h[n]^2},
\end{equation}
where $fs$ denotes the sampling rate.
The resulting $\mathrm{K}$ = 15 parameters are referred to as raw acoustic parameters $a_k$, with $1 \le k \le 15$.

\subsection{Evaluation metrics}
The word error rate (WER) of an ASR engine is calculated as:
\begin{equation}
\mathrm{WER} = \frac{D + S + I}{T},
\label{eq:WER}
\end{equation}
where $T$ is the total number of words, and
$D$, $S$, $I$ are the number of erroneous deletions, substitutions, and insertions, respectively.

To determine performance of the proposed prediction models, we calculate the absolute Pearson correlation coefficient, \ac{RHO}, and the \ac{RMSE} between true and predicted WER.

\subsection{ASR models}
\label{sec:ASR_models}
We use two state-of-the-art ASR models to evaluate the proposed methods to predict WER.
The baseline model was obtained using the Kaldi recipe ``s5'' for LibriSpeech~\cite{Kaldi} to train an ASR model on the clean 100-hour LibriSpeech corpus (\ASR{1}). A more realistic training procedure for evaluating ASR performance on reverberant speech is to provide the network with reverberant training samples. Here we re-train the ASR model on the 100-hour corpus with 50\% of the samples convolved with AIRs drawn randomly from the set described in \cref{sec:corpus} (\ASR{2}).

Recently, Ravanelli et al.\ proposed ASR models based on the Kaldi recipe that achieve state-of-the-art performance for LibriSpeech~\cite{pytorch-kaldi}. We train their proposed model based on Light Gated Recurrent Units (liGRUs) and feature-space Maximum Likelihood Linear Regression (fMLLR) features, both
on the clean 100-hour LibriSpeech corpus (\ASR{3}) as well as the 50\% reverberated corpus described above (\ASR{4}).

\begin{figure*}
    \centering
    \includegraphics{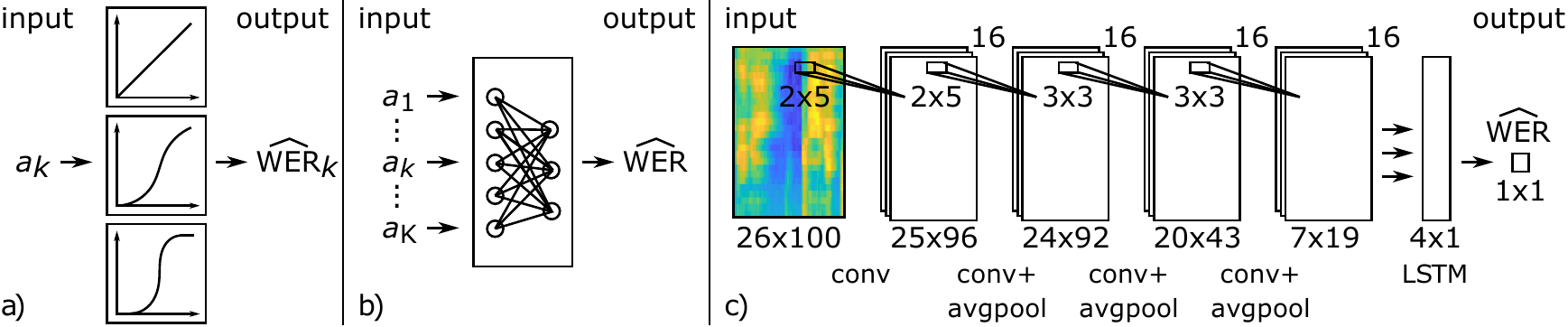}
    \caption{WER prediction models: a) direct fit of raw acoustic parameters, $a_k$, via \cref{eq:polyfit} or \cref{eq:sigfit} and b) neural network fit; c) proposed CNN-LSTM model operating non-intrusively on reverberant speech samples. }
    \label{fig:diagrams}
    \vspace{-0.4cm}
\end{figure*}

\begin{figure}
    \centering
    \includegraphics{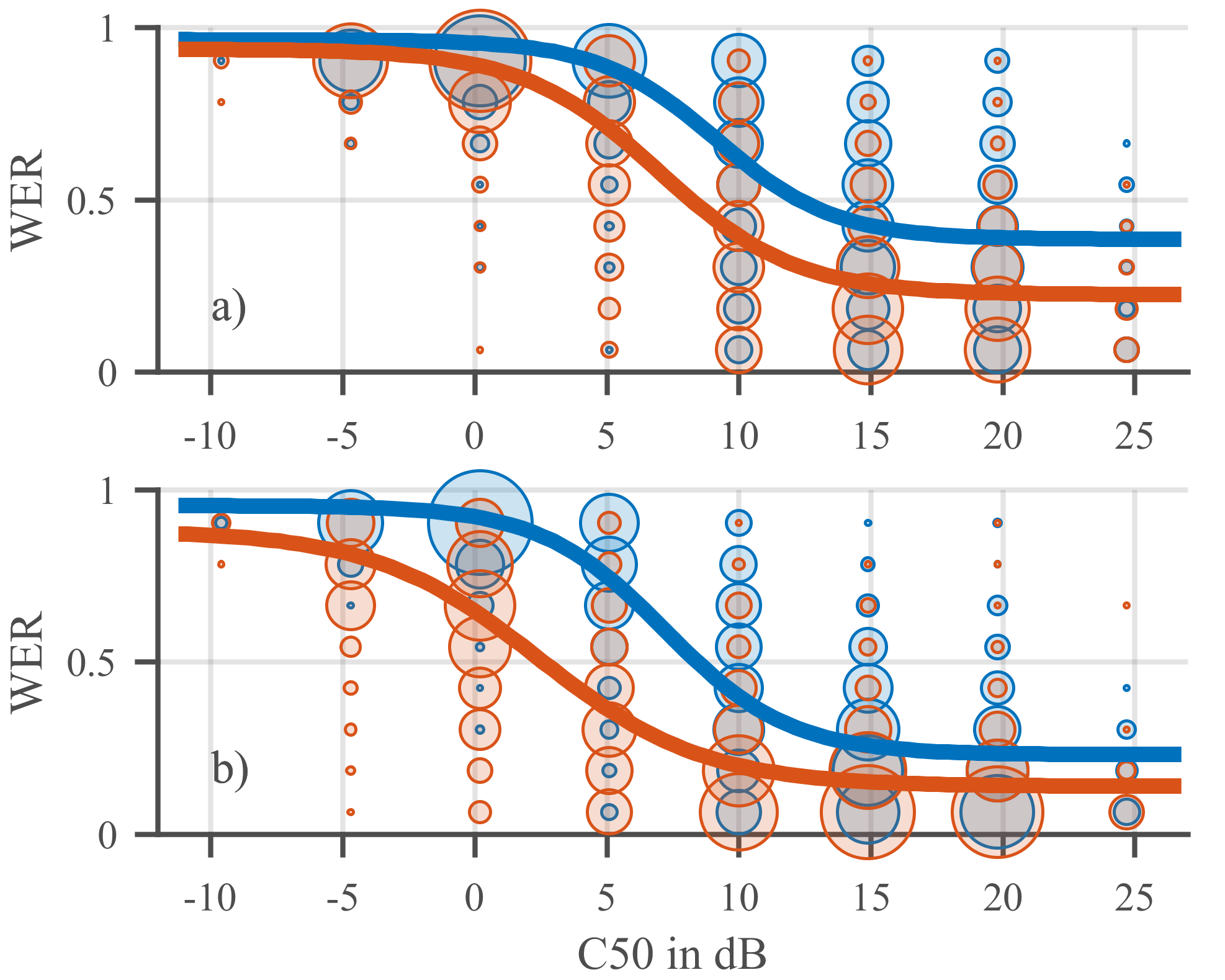}
    \caption{Sigmoidal C50 fit for a) \ASR{1} (blue) and \ASR{2} (orange), and b) \ASR{3} (blue) and \ASR{4} (orange); bubble size indicates sample count.}
    \label{fig:C50_vs_WER}
    \vspace{-0.2cm}
\end{figure}

\section{Predicting WER from acoustic parameters}
\subsection{Polynomial and sigmoidal fit of AIR parameters to WER}
\label{sec:direct_fit}
A number of acoustic parameters have  previously been shown to be correlated with error rates for phoneme recognition tasks. Here we explore their ability to generalize as predictors of WER. The estimated WER is obtained by mapping a raw acoustic parameter, $a_k$ (see \cref{sec:AIR}), to the true WER via a polynomial fit:
\begin{equation}
    \widehat{\mathrm{WER}}_\text{p}(\symvec{p},a_k) = p_{M}a_k^M + p_{M-1}a_k^{M-1} + \cdots + p_{0}a_k^{0},
    \label{eq:polyfit}
\end{equation}
where $\symvec{p} = [p_M, \cdots p_0]$,
and $M \in \{1,3\}$ is the polynomial order. Second-order polynomials are not considered as the mapping of $a_k$ to WER is assumed to be monotonic.

As can be seen in \cref{fig:C50_vs_WER},
the WER as a function of C50 may plateau at the lower end where the ASR model reaches clean speech performance, as well as at the upper end as the WER approaches 100\% (though for corner cases, WER as defined in \cref{eq:WER} may exceed 100\%).
Therefore, we propose a sigmoidal fit \cref{eq:sigfit}, as an alternative to first and third order polynomial fits:
\begin{equation}
    \widehat{\mathrm{WER}}_\text{s}(\symvec{q},a_k) = q_1 + \frac{q_2 - q_1}{1 + \exp\left(q_3(a_k - q_4)\right)}.
    \label{eq:sigfit}
\end{equation}
Coefficients $q_i$ are derived by  squared  error minimization:
\begin{equation}
    \underset{\symvec{q}}{\arg\min}~||\mathrm{WER} - f(\symvec{q},a)||^2_2,
    \label{eq:p}
\end{equation}
where $||\cdot||_2$ is the $L_2$ norm,
and
$f(\symvec{q},a)$ is the fitting function as defined in \cref{eq:polyfit} or \cref{eq:sigfit}. For evaluation, the minimization in \cref{eq:p} is performed on the training set and tested on a hold-out set, to ensure the fit generalizes to unseen data.
The goodness of fit is determined using  the \ac{RMSE} and \ac{RHO} between the true and predicted WER.


\subsection{Neural network fit of AIR parameters to WER}
\label{sec:MLP}
To better understand the interaction of the acoustic parameters and their effect on WER we performed a  \ac{PCA}~\cite{Pearson1901}.
A total of three components explained {97\%} of the variance.
We further studied the combined predictive power of the AIR parameters using a multilayer perceptron (MLP), as illustrated in \cref{fig:diagrams}b. We kept the network complexity low, following \ac{PCA} findings.
{The network consisted of $L$ fully connected hidden layers, with $N_l$ neurons each. }
The estimated parameters of \cref{sec:AIR} comprise the input of the network, with WER as the output.
The evaluation set described in \cref{sec:corpus} was used for training, validating and testing the model.
The network parameters were optimized on the validation set via grid search and  mean-square-error loss, over 30 epochs, {with $L \in [0, 4]$ and $N_l \in [1, 32]$, }
a 5\% drop-out rate, and rectified linear unit (ReLU) activation functions. We propose a  network with one fully connected layer with 3 neurons.
\section{Predicting WER blindly from reverberant speech using a CNN-LSTM model}
In practice, the clean speech or raw AIRs of reverberant speech may not be available, e.g., if the samples stem from actual device recordings. We propose an alternative, data-driven approach to predict WER directly from the reverberant speech samples. The assumption is that much like blind or non-intrusive acoustic parameter estimation can be used as a proxy for estimating ASR performance~\cite{Parada2016}, a neural network model can be trained to extract features from reverberant speech that are correlated with WER. The proposed method assumes reverberant speech samples transcribed by an ASR engine and the corresponding WER per utterance calculated by \cref{eq:WER}. The same data split as described in \cref{sec:corpus} is used. A neural network model is trained on the TR set to predict the WER non-intrusively, using a squared-error loss function. The CV set is used to monitor training progress and ensure a good model fit.

The goal of the proposed model is to predict WER on the unseen test data TE, using a neural network that is substantially more light-weight than a fully-blown, state-of-the-art ASR model.
The input features are inspired by approaches used for ASR, i.e., the proposed model may lend itself to operating directly on the features extracted by the ASR engine, thus saving computations. The samples are processed in frames of 640 samples, corresponding to a frame length of 25~ms at a sampling rate of 16~kHz, with a hop size of 160 samples, i.e., 10~ms. 26 Mel-frequency bins are extracted per frame, and 100 consecutive frames are combined to a $26 \times 100$ feature matrix, which corresponds to about 1~s of speech. We propose a 4-layer convolutional neural network (CNN) model with ReLU activation functions that operates on the input feature matrices of each utterance. All but the first layer are followed by $3 \times 3$ average-pooling layers with a stride of (2,2). The CNN is followed by a single long short-term memory (LSTM) layer with four cells that accumulates frame-level estimates to produce one WER prediction per utterance. The \numprint{7809} model parameters are trained over 100 epochs in about 4 hours on 4 GPUs. \Cref{fig:diagrams}c illustrates the proposed CNN-LSTM model.

\begin{table}[]
\setlength{\tabcolsep}{4pt}
    \centering
    \caption{ASR performance for clean and reverberant speech. 
    }
    \label{tab:ASR_results}
    \begin{tabular}{l S[table-format=2.2] S[table-format=2.2] S[table-format=2.2] S[table-format=2.2]}
    \toprule
      & \ASR{1} & \ASR{2} & \ASR{3} & \ASR{4}\\
    \cmidrule{2-5}
    WER, clean speech [\%]       & 12.71 & 13.30 & 8.66 & 9.30 \\
    WER, reverberant speech [\%] & 69.22 & 54.76 & 56.22 & 35.17\\
    \bottomrule
    \end{tabular}
\end{table}

\begin{figure}
    \centering
    \includegraphics{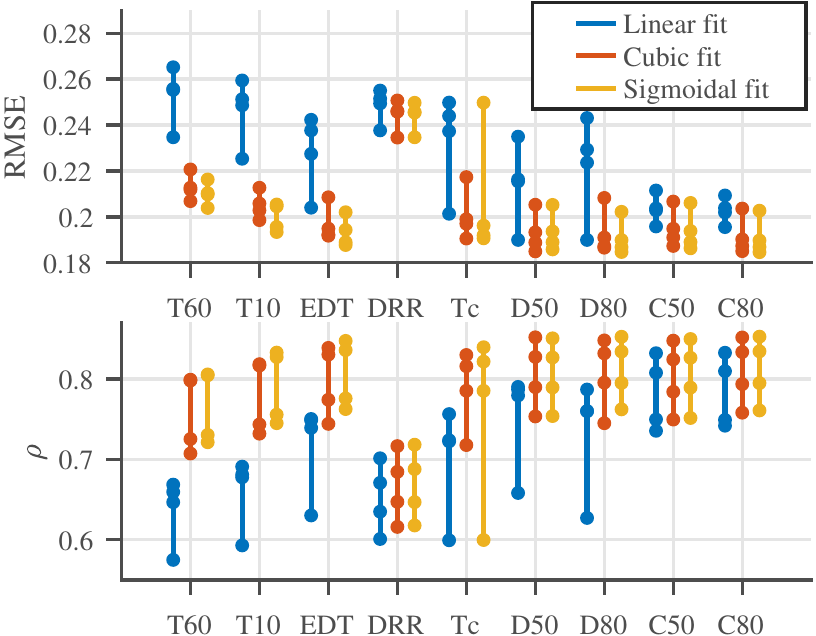}
    \caption{\ac{RMSE} (top) and \ac{RHO} (bottom) of the WER fit based on individual AIR parameters, $a_k$ (dots show results for the four ASR models). The number of parameters shown is reduced for clarity of presentation. Note that $\mathrm{D}_{t}$ and $\mathrm{C}_{t}$ parameters have a negative correlation.}
    \label{fig:AIR_WER}
    \vspace{-0.2cm}
\end{figure}

\section{Experimental evaluation}
Experiments are performed on four ASR models, described in \cref{sec:ASR_models}, and using the evaluation set described in \cref{sec:corpus}.

Performance of the four ASR models is summarized in \cref{tab:ASR_results}, in terms of WER for the clean LibriSpeech test set as well as the reverberant evaluation set (TR + CV + TE, as described in \cref{sec:corpus}, i.e., note that TR refers to the training set used to train the WER predictors, not the ASR model). As can be seen, the WER increases dramatically for reverberant speech for all four models. However, the models trained on a set containing 50\% reverberant samples, i.e., \ASR{2} and \ASR{4}, clearly outperform the models trained entirely on clean speech, at the cost of a slight decrease in clean-speech performance. \cref{fig:C50_vs_WER} clearly shows this effect as well.
We assume that these WERs are representative of the performance of state-of-the-art ASR models in challenging reverberant conditions.

\Cref{fig:AIR_WER} illustrates the goodness of fit for the acoustic parameters extracted from the AIR (see \cref{sec:AIR}), for the four tested ASR models and the TE set. Clarity, C$_t$, definition, D$_t$, and center time, Tc, best predict the WER. 
For the linear fit, C50 exhibits slightly lower correlation with WER, \ac{RHO} $= 0.78$ averaged across ASR models, than the correlation reported by Parada et al.\ with phoneme error rate~\cite{Parada2016}. This seems to confirm their hypothesis that phoneme error rate better reflects the effect of reverberation, as it eliminates the effect of the language model~\cite{Parada2016}. A third-order polynomial or sigmoidal fit, with coefficients derived from the TR set, improved the predictive power of all tested AIR parameters on the TE set.

\begin{table}[]
    \begin{center}
    \setlength{\tabcolsep}{3pt}
    \caption{WER prediction results.}
    \label{tab:Results}
    \begin{tabular}{l S[table-format=1.2] S[table-format=1.2] S[table-format=1.2] S[table-format=1.2] S[table-format=1.2] S[table-format=1.2] S[table-format=1.2] S[table-format=1.2]}
    \toprule
     & \multicolumn{4}{c}{\ac{RMSE}} & \multicolumn{4}{c}{\ac{RHO}}\\
     & {\ASR{1}}& {\ASR{2}}& {\ASR{3}}& {\ASR{4}}& {\ASR{1}}& {\ASR{2}}& {\ASR{3}}& {\ASR{4}} \\
    \cmidrule(lr){2-5}
    \cmidrule(lr){6-9}

     C50, linear   & 0.21   & 0.20   & 0.20   & 0.20   &   0.74   & 0.81   & 0.83   & 0.75     \\
      C50, cubic   & 0.21   & 0.19   & 0.19   & 0.19   &   0.75   & 0.82   & 0.85   & 0.78     \\
    C50, sigmoid   & 0.21   & 0.19   & 0.19   & 0.19   &   0.75   & 0.83   & 0.85   & 0.79     \\
    MLP            & 0.20   & 0.18   & 0.18   & 0.18   &   0.77   & 0.85   & 0.86   & 0.80  \\

    \cmidrule(lr){2-5}
    \cmidrule(lr){6-9}


CNN-LSTM$^\ast$ & 0.20   & 0.18   & 0.18   & 0.18   &   0.77   & 0.85   & 0.86   & 0.81\\


    \bottomrule
    \end{tabular}
    \end{center}
\vspace{-0.2cm}
{\scriptsize $^\ast$Blind WER estimation, i.e., without knowledge of AIR parameters or clean reference.}
\vspace{-0.4cm}
\end{table}

\Cref{tab:Results} summarizes the WER prediction results. In line with prior findings, C50 serves as a relatively good predictor for WER for the tested state-of-the-art ASR models. Applying a third-order polynomial or sigmoidal fit to the raw estimates, using coefficients derived from the TR set (cf.\ \cref{sec:direct_fit}), improves accuracy in all cases. The proposed MLP provides more accurate WER estimates (in terms or \ac{RMSE} and \ac{RHO}) than a direct fit.
Finally, the proposed CNN-LSTM model, which estimates WER blindly from reverberant speech samples, slightly outperforms all other tested estimators. Note that while a one-way Anova does not indicate a statistically significant reduction in error rate compared to a linear C50 fit, the improvement is consistent across all tested ASR models and obtained without reference or knowledge of the AIR parameters.

\section{Conclusion}
Our results indicate that word error rate (WER) of an automatic speech recognition (ASR) model processing reverberant speech can be predicted directly from acoustic impulse response (AIR) parameters, as well as blindly from the reverberant utterances. We tested two state-of-the-art ASR models, trained either entirely on clean speech or on a combination of clean and reverberant speech. Our results are in line with prior art, indicating that clarity, i.e., C50 and C80, as well as definition, i.e., D50 and D80, are highly correlated with WER. We show that WER prediction can be improved by using a third-order polynomial or sigmoidal fit or a neural network to map AIR parameters directly to WER. 
Finally, we propose a convolutional neural network (CNN) model that predicts WER blindly from the reverberant speech samples, without knowledge of the underlying AIR parameters or access to the clean speech. The proposed models may prove useful for developing speech enhancement models or unsupervised ASR (pre-)training. A noteworthy limitation of the present work is that noise is not considered. Future work includes extending the WER estimators to reverberant speech in noise as well as studying how well they generalize to unseen ASR models.

\section{Acknowledgments}
The authors thank Yifan Gong from Microsoft Speech Services for advice on the speech data and ASR models.
%


\bibliographystyle{IEEEtran}
\bibliography{references.bib,IEEEsettings.bib}
%
%
%
%
%
%
%
%
%

\end{sloppy}
\end{document}